# Gate-tunable graphene-based Hall sensors on flexible substrates with increased sensitivity

Burkay Uzlu [1,2*], Zhenxing Wang[1], Sebastian Lukas[1,2], Martin Otto[1], Max C. Lemme[1,2] and Daniel Neumaier[1]

[1]Advanced Microelectronic Center Aachen (AMICA), AMO GmbH, 52074 Aachen, Germany

[2]Chair of Electronic Devices, RWTH Aachen University, 52074 Aachen, Germany

[*]Author to whom correspondence should be addressed. Email: uzlu@amo.de

**Abstract**

We demonstrate a novel concept for operating graphene-based Hall sensors using an alternating current (AC) modulated gate voltage, which provides three important advantages compared to Hall sensors under static operation: 1) The sensor sensitivity can be doubled by utilizing both n- and p-type conductance. 2) A static magnetic field can be read out at frequencies in the kHz range, where the 1/f noise is lower compared to the static case. 3) The off-set voltage in the Hall signal can be reduced. This significantly increases the signal-to-noise ratio compared to Hall sensors without a gate electrode. A minimal detectable magnetic field $B_{min}$ down to 290 nT/√Hz and sensitivity up to 0.55 V/VT was found for Hall sensors fabricated on flexible foil. This clearly outperforms state-of-the-art flexible Hall sensors and is comparable to the values obtained by the best rigid III/V semiconductor Hall sensors.

**Introduction**

Magnetic field sensors are widely used in several key industries such as consumer electronics, automotive, healthcare and robotics[1,2], where they provide position precision, speed detection, switching applications or current monitoring[3]. Hall effect sensors currently dominate the market in terms of these operations[4–8] and the key figures of merit for these sensors are their magnetic field resolution ($B_{min}$) and current ($S_i$) and voltage related ($S_v$) sensitivities[9]. Today's technology utilizes mainly silicon based Hall sensors because of the advanced silicon CMOS technology that enables reliable and cost efficient production[10]. However, applications of silicon based Hall sensors are limited by their relatively low sensitivity when compared to Hall sensors based on III/V semiconductors such as InAs and GaAs[11–13], and by their mechanical stiffness. Graphene, because of its very high mobility μ and low charge carrier density n ($S_v \sim \mu$ and $S_i \sim 1/n$) as well as its mechanical flexibility, appears to be ideally suited as the sensing material in Hall effect sensors[14–21]. Previously, graphene based Hall sensors were demonstrated, that outperformed silicon and other III/V semiconductors based rigid sensors[8,9,15,22–24]. In addition, graphene-based Hall sensors already significantly outperform all other technologies on flexible substrate[22]. However, the sensitivity of flexible graphene Hall sensors is not yet comparable to rigid Hall sensors based on conventional semiconductors, at least if the graphene Hall sensors are fabricated with a scalable approach.

In this work, we introduce a novel concept for increasing the signal-to-noise ratio in graphene-based Hall sensors with an AC modulated gate electrode, on flexible substrate with local top-gate. Such dynamic gating provides three advantages compared to graphene Hall sensors under static operation: First, the AC modulated gate voltage allows reaching sensitivity maxima for n- *and* p-type conductance, which doubles the effective sensitivity. Second, a static magnetic field can be read out at higher frequencies, where 1/f noise is significantly lower. And third, offsets in the Hall voltage can be reduced by adjusting a static voltage superimposed on the AC voltage. The basic operation mechanism of such a top-gated Hall sensor is illustrated in Figure 1a. While the second and third advantage can be achieved by other methods, *e.g*. AC modulation of the supply voltage for the former or spinning current readout for the latter[25], there exists no approach offering all three advantages. In the following, we describe the fabrication and detailed characterization of top-gated graphene Hall sensors on flexible substrate and demonstrate an increased signal-to-noise ratio. These Hall sensors do not only set a new benchmark for flexible Hall-sensors, but are also on a par with the best III-V based rigid Hall sensors.

**Results and Discussion**

Schematics and the optical microscope image of the fabricated Hall sensors on polyamide (PI) substrate are shown in Figure 1b and Figure 1c. The transfer characteristic of one representative device is shown in Figure 1d for both sweeping directions from -4 V to 4 V and back. These field effect transistor measurements have been conducted to check the integrity and quality of the devices. The transfer characteristic shows little hysteresis, which demonstrates successful passivation by the $Al_2O_3$[26], and also a low residual doping level of the graphene of $3.5 \cdot 10^{11}$ cm$^2$. The two-probe field effect mobility of the device is 2830 cm$^2$/Vs, including the contact and access resistance. Detailed device statistics of 51 out of total 54 devices fabricated on the same chip are shown in the Supplementary Information and demonstrate good reproducibility. Three devices which did not function are excluded from the statistics.

First, conventional Hall effect measurements have been performed with constant gate voltage in order to extract the basic parameters of the device. The measurement configuration is shown in Figure 1b. A constant bias voltage ($V_C$) and a magnetic field perpendicular to the graphene channel, ranging from 7.2 to 28.8 mT, has been applied and the current ($I_C$) and the Hall voltage ($V_H$) have been measured. In addition, the charge carrier concentration of the graphene, and thus the sensitivity of the device, has been controlled by applying a gate voltage ($V_G$). The normalized Hall voltage $\Delta V_H$, which represents the measured Hall voltage minus the offset voltage at zero magnetic field ($\Delta V_H = V_H - V_{H,\,B=0}$) is plotted as a function of gate voltage in Figure 2a with a constant $V_C = 300$ mV. The voltage ($S_v$) and current related ($S_i$) sensitivities can be derived from this measurement using equations[15] (1) and (2):

$$S_v = \frac{1}{V_c}\left|\frac{\partial V_H}{\partial B}\right| \qquad (1)$$

and

$$S_i = \frac{1}{I_c}\left|\frac{\partial V_H}{\partial B}\right| \qquad (2)$$

Figures 2c and 2d show $S_v$ and $S_i$ as a function of top gate voltage at a bias voltage of $V_C = 300$ mV. The sensitivities crucially depend on the doping level of the graphene with a slight asymmetry between hole and electron transport regime. The maximum values for the sensitivities $S_i$ and $S_v$ are found to be 800 V/AT and 0.278 V/VT, respectively, for this device. We further observe a linear dependence of $\Delta V_H$ regarding $V_C$ (Figure 2b), which indicates that the sensitivities are independent of the applied bias and the sensors can be operated with low power. The corresponding charge carrier mobility ($\mu$) can be calculated according to equation[9] (3):

$$\mu = S_v \frac{L}{W} \qquad (3)$$

where L and W are the length and width of the graphene channel in the fabricated Hall sensor device. This results in a charge carrier mobility of 4400 cm$^2$/Vs at the $S_v$ maxima, which turned out to be an average value across the chip. Detailed performance analyses of all devices on the chip are shown in Figures S1 and S2 in the Supplementary Information. The device with the highest values on this chip showed $S_V$ = 0.35 V/VT, corresponding to a mobility of 5600 cm²/Vs.

The characteristics of the Hall sensors under AC gate modulation have been measured in a different setup, illustrated in Figure 3a. Here, we report data measured on the same device as discussed in Figire 2. In this setup, a signal generator is used to modulate the gate voltage, which consists of a static offset voltage plus an AC voltage. An SRS 380 lock-in amplifier is used to demodulate the read-out signal ($V_H$). The modulation frequency is fixed at 2 kHz for all measurements, which is well below the calculated RC bandwidth of the Hall sensor of 1.7 MHz. Figure 3b shows the $\Delta V_H$ as a function of the peak-to-peak AC voltage. In this measurement a constant $V_C$ of 300mV was applied. As expected from the gate voltage dependent sensitivity, the Hall signal increases first with increasing AC amplitude and decreases again after reaching a maximum at ~1.5 V. To directly compare the Hall voltage under DC and AC operation, 45 second long measurements have been performed while the magnetic field was alternated between zero and values up to 28.8 mT (Figure 3c). In both measurements, $V_C$ = 300 mV. The gate voltages have been set to achieve maximum sensitivity, i.e. $V_G$ = -1.2 V for the DC measurement and $V_G$ = 1.5 V AC voltage in amplitude for the AC case. The effective Hall signal and thus the sensitivity is approximately a factor 2 higher for the AC case, confirming the advantage of the AC gate modulation. The extracted sensitivity $S_V$ is 0.55 V/VT and 0.278 V/VT for the AC and

DC case, respectively (Supplementary Information Figure S3). In addition, the noise in the AC measurements is significantly lower compared to the DC case. However, we note that this noise reduction is mainly due to the sensor read-out using a lock-in amplifier for the AC case, which is not possible under static operation.

The noise power spectral density of the sensor is shown in Figure 4a, confirming that the major source of noise has a 1/f dependency for the measured frequency range up to 12 kHz. Hence, higher read-out frequencies lead to lower noise levels. The magnetic resolution $B_{min}$ represents another key parameter for Hall sensors in addition to the current and voltage related sensitivities. It can be calculated using the noise spectral density and the sensitivity of the sensor using equation[15] (4):

$$B_{min} = \frac{\sqrt{P_v}}{(S_v \cdot V_c)} \qquad (4)$$

The magnetic resolution is derived from $P_V$ (Figure 4b). At 2 kHz, $B_{min}$ is found to be 500 nT/$\sqrt{Hz}$. The lowest value measured on the chip was 290 nT/$\sqrt{Hz}$. Comparison of our findings with literature values for the state-of-the-art Hall sensor elements made from silicon, graphene and other III/V based semiconductors on rigid and flexible substrates are shown in Table I. The table clearly shows that our CVD graphene based Hall sensors on flexible substrate remarkably outperforms all the other Hall sensor elements on flexible substrate and are highly competitive with respect to all existing technologies on rigid substrate. The measured minimum magnetic resolution of our Hall sensors also outperform the state-of-the-art Hall elements based on Si and is close to the very best values achieved by AlInSb[4,6] and exfoliated graphene[9] based Hall sensors. It should be noted that the highest[9,22] values shown in the Table I, have been achieved by non-scalable micro-mechanical exfoliated graphene and hBN and/or in vacuum conditions.

Finally, the stability of the devices for flexible applications has been investigated. Bending tests have been performed where the device was exposed to varying strain values. Figure 5a shows a photograph of a flexible chip after it has been peeled off mechanically from the silicon carrier substrate. For electrical measurements, the PI substrate has been bent under different bending radii of 25.4 mm, 12.7 mm, and 6.4 mm, respectively (Figure 5b inset). Bending has been performed one time for each radius, with the sensor on the outer side of the bending curvature (tensile strain). Hall measurements have been performed afterwards in flat status in constant voltage mode. In addition to the measurements with different bending radii, the sensors have undergone up to 1000 bending cycles with a bending radius of 6.4 mm (Figure 5b). No major degradation of the device performance is observed up to 1000 bending cycles. These results expand the application space of hall sensors to flexible electronics, *e.g.*, wearable sensors for personal fitness or healthcare systems.

**Conclusion**

In summary, we have introduced the concept of tunable graphene Hall sensors through AC modulated gate voltages and performed an in-depth characterization of the devices. The AC top gate modulation increased the Hall signal, and thus the sensitivity, by a factor of ~2 compared to Hall sensors under DC operation. In addition, AC gate modulation enables read out of the sensors at kHz frequencies, where the noise level is significantly (by a factor of ~$10^5$) lower compared to the DC case and thus allows detecting much smaller magnetic fields. In addition, the device concept and the fabrication process are compatible with large-scale flexible substrates. The performance does not degrade under bending and strain, which enables highly sensitive flexible Hall sensors. These results set a new benchmark for flexible Hall-sensors.

## Methods

Graphene-based Hall sensors are fabricated on spin-coated PI substrate with conventional photolithography technology. First, the flexible substrate is prepared by spin-coating PI in liquid form on a Silicon handling substrate. After curing, the resulting thickness of the solidified PI film is about 8 μm. During the entire fabrication process, Si is used as a supportive layer for the PI. After the PI substrate preparation, commercially available, large area, chemical vapor deposited graphene is transferred by a PMMA assisted method[27]. Graphene is patterned by oxygen plasma, and contacts to graphene are fabricated by sputtering 50 nm Ni, followed by a lift-off process. The top gate dielectric consists of 40 nm $Al_2O_3$ deposited by atomic layer deposition and the top-gate electrode is fabricated by sputter deposition of 20/500 nm Ti/Al and lift-off. Wet buffered oxide etchant is used to open vias in the $Al_2O_3$ to access the contacts. All measurements have been performed under ambient condition at room temperature. The devices have been measured in flat status with the PI still on silicon, but the entire flexible stack has been peeled off mechanically from the silicon carrier substrate for electric characterization under bending.

**Acknowledgements**

This work was financially supported by the European Commission under the project Graphene Flagship (contract no. 785219) and by the German Science Foundation (DFG) within the priority program FFlexCom Project "GLECS" (contract no. NE1633/3).

**Author Contributions**

B.U. and D.N. proposed the idea and planned the experiments. M.O. helped for sample preparation. B.U. fabricated the devices and performed all the experiments. Z.W. and S.L. helped for the measurements. B.U analyzed the data and wrote the manuscript. The manuscript was critically reviewed by M.C.L. and D.N. All authors reviewed the manuscript and approved the final version of the manuscript.

**Additional Information**

**Supplementary Information**

Detailed statistics the devices fabricated on the same chip, Hall mobility, field effect mobility and charge carrier density measurements of graphene, Hall effect measurements with DC and AC modulated gate voltage and comparison of different high performance Hall sensors.

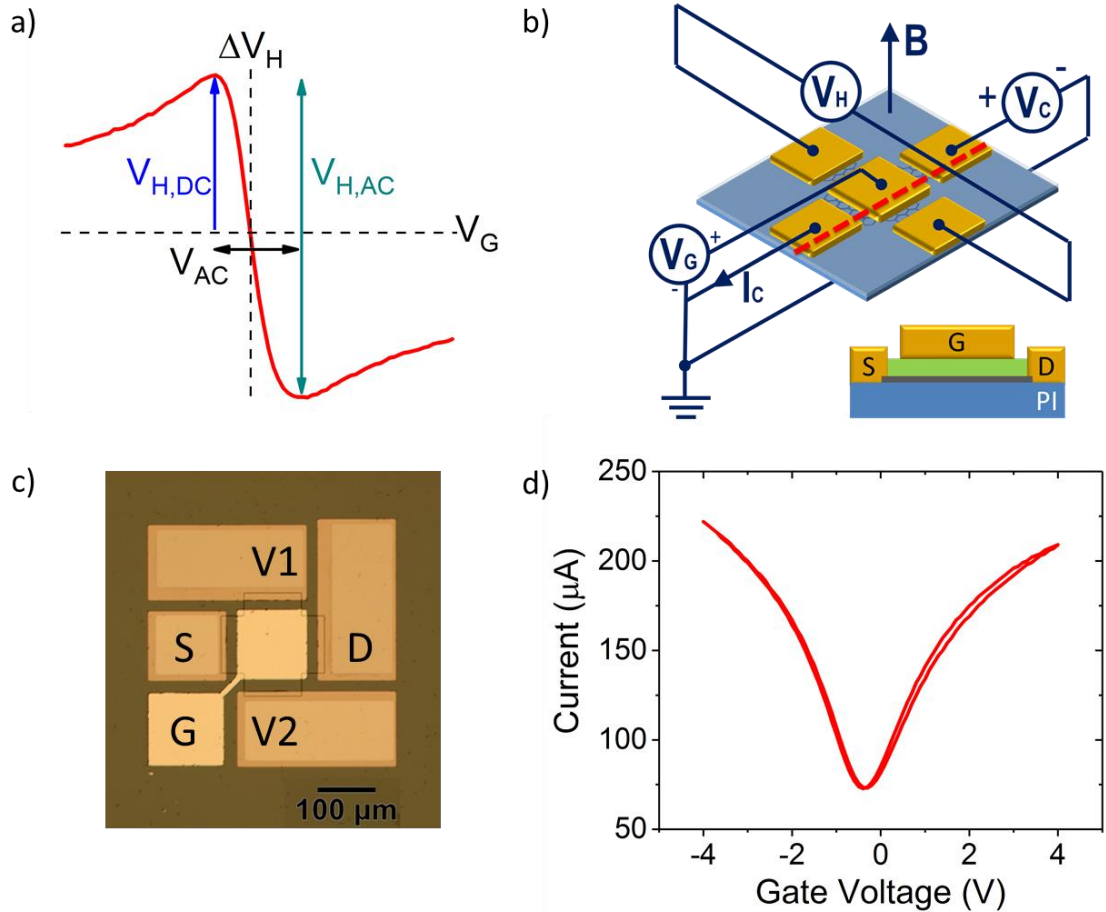

**Figure 1:** Device schematic and the IV characteristics of the top gated graphene based flexible hall sensor. a) Comparison of measured $\Delta V_H$ and basic operating principles using a DC gate voltage and an AC modulated gate voltage across the charge neutrality point. In the latter, both sensitivity maxima for n- and p-type conductance can be utilized and the effective sensitivity is doubled by AC gate modulation. b) Isometric device schematic of the top gated graphene hall sensor with corresponding biasing scheme (top). Schematic illustration of the cross section along the red dashed line in the device schematic (bottom). Graphene and $Al_2O_3$ encapsulation are indicated in grey and green colors respectively. c) Optical microscope image of a device after fabrication with four probing metal pads and top gate. A constant bias voltage $V_C$ is applied between the contacts S and D and $\Delta V_H$ is measured between the contact V1 and V2. Gate voltage

$V_G$ is applied to the contact G. d) Room temperature two terminal top gate characteristic of the fabricated device at $V_C$ of 300mV.

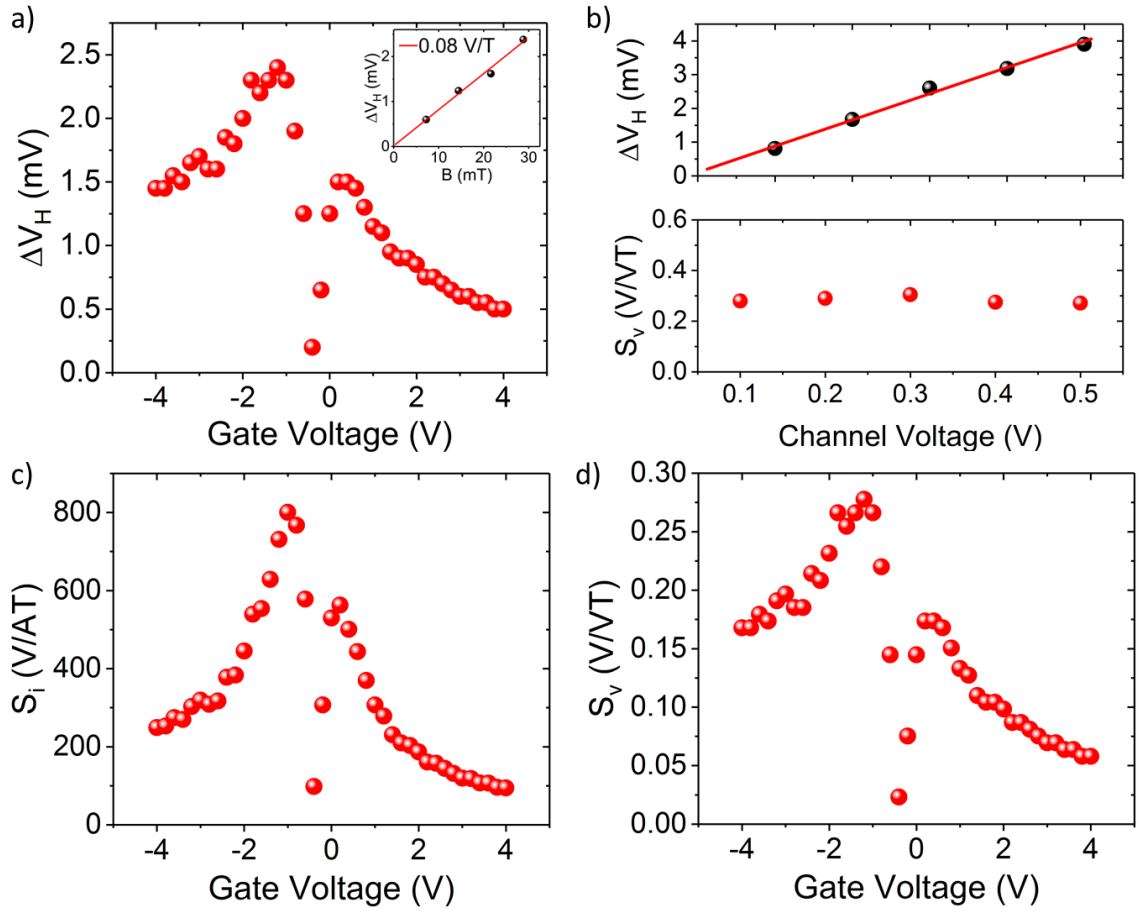

**Figure 2:** Hall measurements of the sensor. a) Normalized Hall voltage $\Delta V_H$ as a function of the gate voltage at $V_C = 300$ mV. The inset shows $\Delta V_H$ as a function of the magnetic field at $V_G = -1.2$ V. b) $\Delta V_H$ and voltage sensitivity $S_V$ as a function of channel voltage $V_C$. c) and d) Absolute values of current sensitivity $S_i$ and $S_v$ plotted against $V_G$ at $V_C = 300$ mV.

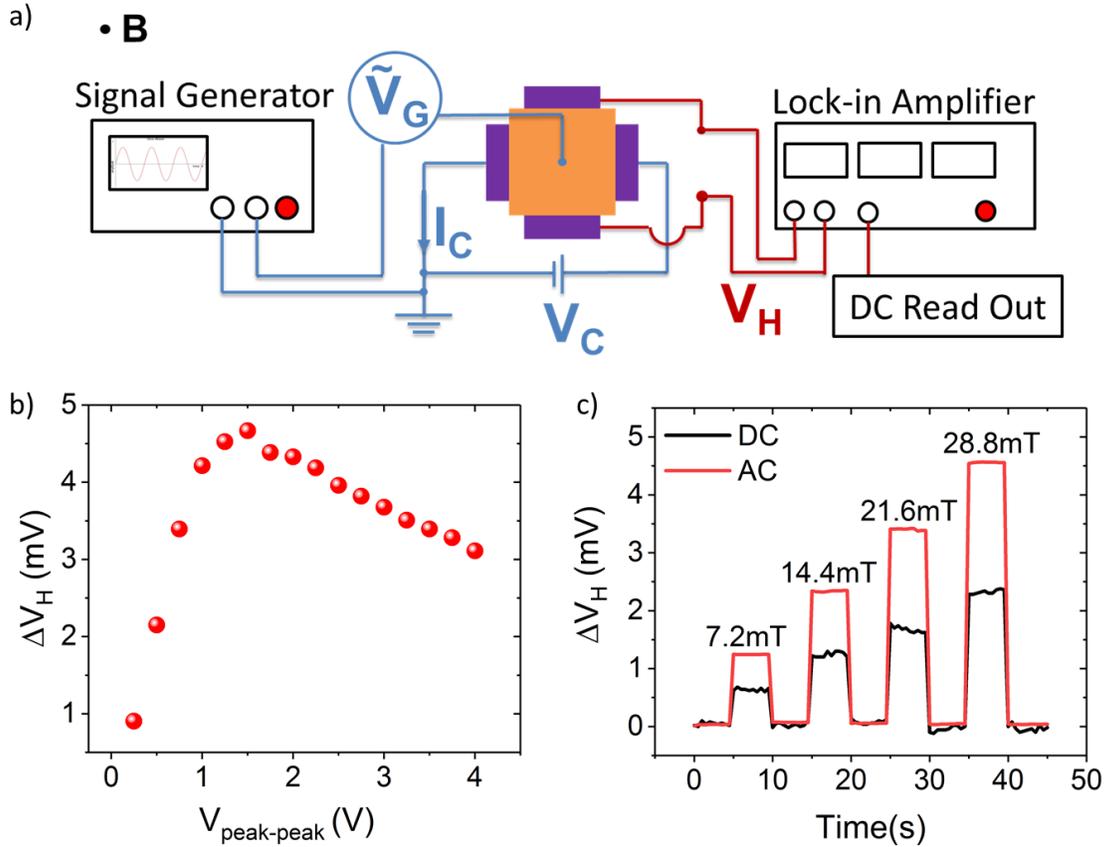

**Figure 3:** Hall measurements of the sensor with gate voltage modulation. a) Illustration of AC gate modulation setup. A signal generator is used to modulate the gate voltage and an SRS 380 lock-in amplifier is used to demodulate the read-out signal ($V_H$). b) Hall voltage response of the device to the varying magnetic field at a peak to peak gate modulation amplitude of 1.5 V c) Hall voltage under DC (black) and AC (red) operation over time, while the magnetic field was stepped alternated between and up to 28.8 mT. For both measurements, $V_C$ was 300 mV and the gate voltage was tuned to maximum sensitivity, i.e. $V_G$ = -1.2 V for the DC measurement and $V_G$ = 1.5 V AC voltage for the AC case.

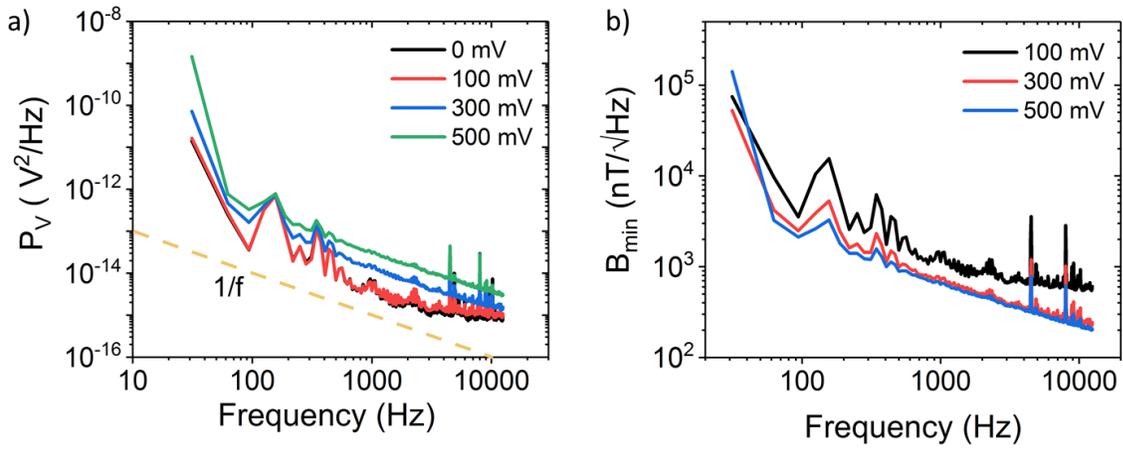

**Figure 4:** Noise and B$_{min}$ measurements. a) Noise power spectral density (P$_v$) as a function of frequency. The dashed line indicates 1/f behavior of the noise. b) Derived magnetic resolution B$_{min}$ of the hall sensor as a function of frequency.

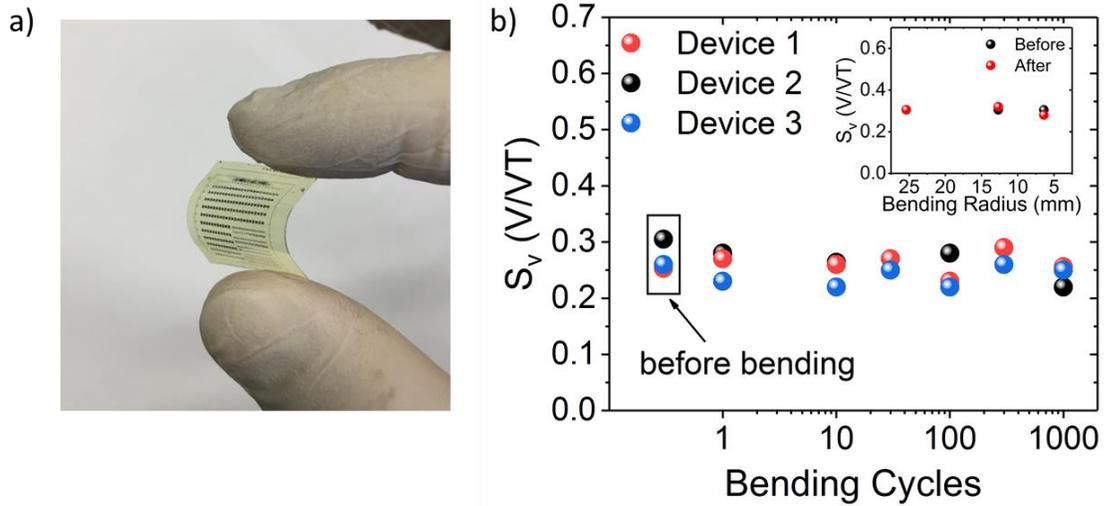

**Figure 5:** Bending tests for flexible Hall sensor. (a) Optical photograph of the flexible chip after it has been peeled off from Si substrate. (b) Bending cycle dependent measurement of the Hall sensor sensitivity. The PI substrate has been bent under bending radii of 6.4 mm up to 1000 times. The inset shows the measured $S_v$ before and after bending tests with different bending radii of 25.4 mm, 12.7 mm, and 6.4 mm. The measurements have been carried out in flat status before and after bending at constant $V_C$ of 300mV and $V_G$ of -1.2V.

**Table 1:** Metrics comparison of different high-performance Hall elements working at room temperature. ('Gr' stands for graphene).

| | Substrate | $S_i$ (V/AT) | $S_v$ (V/VT) | $B_{min}$ (nT/$\sqrt{Hz}$) | Frequency (kHz) | Conditions |
|---|---|---|---|---|---|---|
| Si[1,28] | Rigid | 100 | 0.1 | 5000 | 3 | n/a |
| AlInSb[4,6] | Rigid | 2750 | 2.2 | 58 | 1 | Vacuum |
| GaAs[28] | Rigid | 1100 | n/a | 800 | 3 | n/a |
| Exfoliated Gr-hBN[9] | Rigid | 4100 | 2.16 | 50 | 3 | Vacuum |
| CVD Gr[29] | Rigid | 800 | n/a | 500 | 3 | Vacuum |
| CVD Gr[30] | Rigid | 2093 | 0.35 | 100 | 3 | Air |
| CVD Gr[8] | Flexible | 75 | 0.093 | n/a | n/a | Air |
| CVD Gr-hBN[22] | Flexible | 2270 | 0.68 | n/a | n/a | Air |
| Bismuth[31] | Flexible | 2.3 | n/a | n/a | n/a | Air |
| This work | Flexible | 1500 | 0.55 | 500 | 2 | Air |
| This work (max) | Flexible | 2580 | 0.68 | 290 | 2 | Air |

# Gate-tunable graphene-based Hall sensors on flexible substrates with increased sensitivity

Burkay Uzlu [1,2*], Zhenxing Wang[1], Sebastian Lukas[1,2], Martin Otto[1], Max C. Lemme[1,2] and Daniel Neumaier[1]

[1]Advanced Microelectronic Center Aachen (AMICA), AMO GmbH, 52074 Aachen, Germany

[2]Chair of Electronic Devices, RWTH Aachen University, 52074 Aachen, Germany

[*]Author to whom correspondence should be addressed. Email: uzlu@amo.de

**Supplementary Information**

**Figure S1**: Measurements of different Hall sensors on the same sample chip

**Figure S2**: Graphene mobility and charge carrier density measurements

**Figure S3**: $S_i$ and $S_v$ measurements with gate modulation

**Figure S4**: AC/DC gated Hall measurements to compare offset values

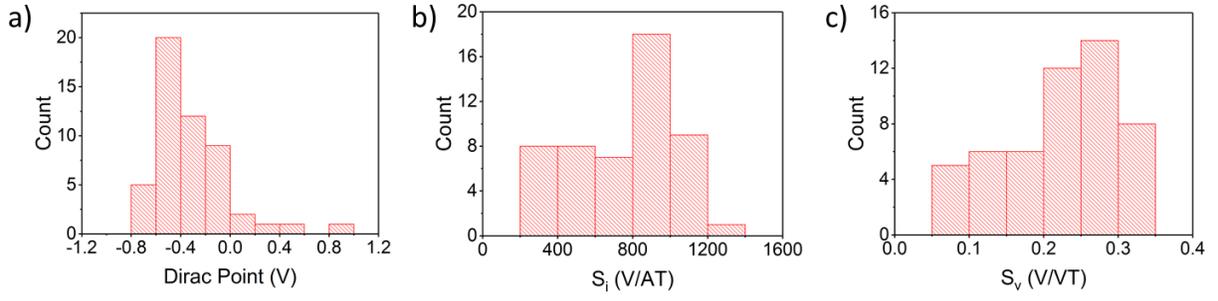

**Figure S1:** Measurements of different Hall sensors on the same sample chip. Histogram of the (a) Dirac point distribution (b) current normalized sensitivity and (c) voltage normalized sensitivity for different Hall sensors on the same sample chip

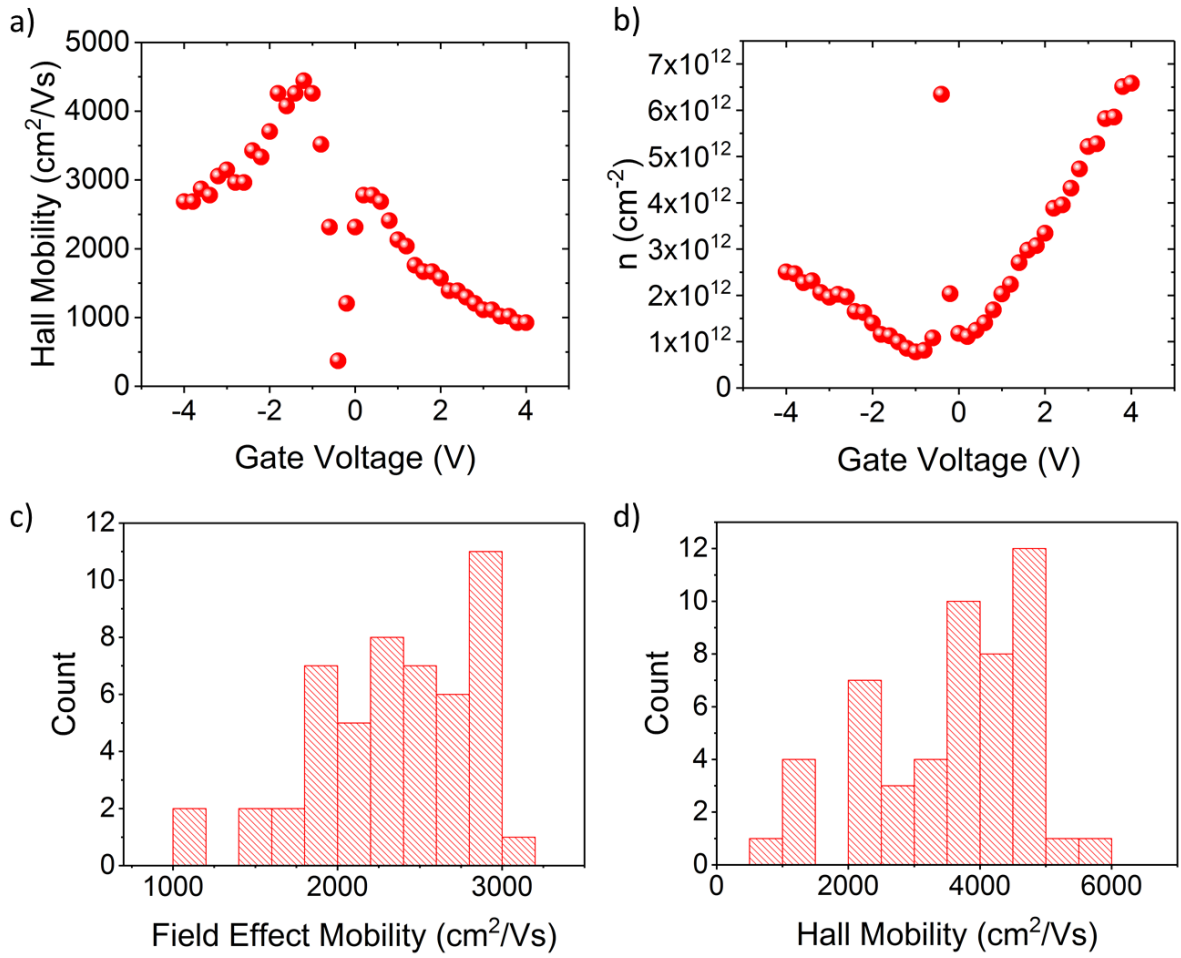

**Figure S2:** Graphene mobility and charge carrier density measurements. a) Hall mobility and b) charge carrier density against $V_G$ at constant $V_G$ of 300mV. Maximum c) field effect and d) hall mobility values for different Hall sensors on the same sample chip.

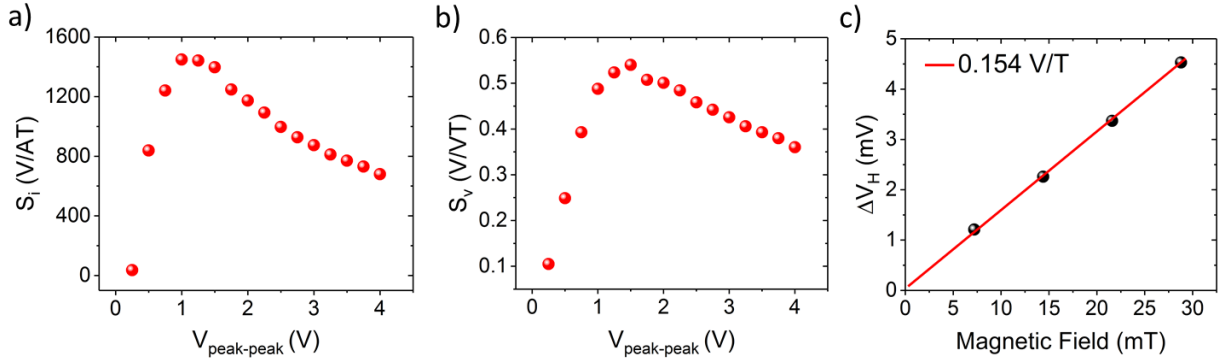

**Figure S3:** Hall measurements of the sensor with gate voltage modulation. a) and b) Absolute values of $S_I$ and $S_v$ plotted against varying gate voltage modulation amplitude at $V_C=300mV$. c) Linear dependence of $\Delta V_H$ to the magnetic field at peak-to-peak gate modulation amplitude of 1.5V.

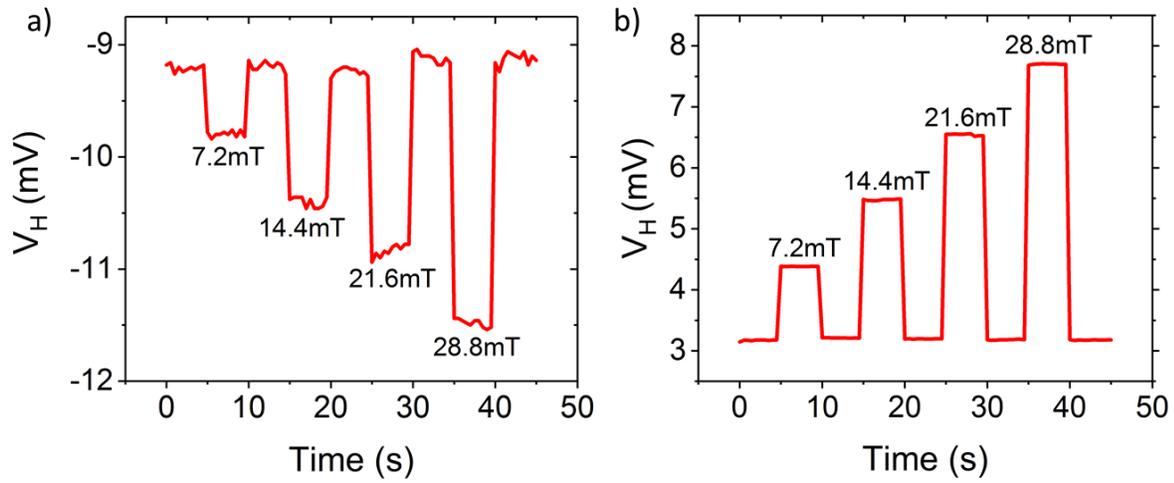

**Figure S4:** The measured hall voltage (with offset) under DC and AC gate voltage operation. A time sweep was performed and the magnetic field was stepped from zero to 28.8 mT. a) Hall voltage with DC gate voltage at $V_G$= -1.2V b) Hall voltage with AC gate voltage operation at peak to peak gate modulation amplitude of 1.5V. $V_C$ is 300mV for both measurements. Offset reduction and doubling of $\Delta V_H$ can be seen from the comparison of two graphs.